# Ageing under stress and mechanical fragility of soft solids of laponite

G. Ranjith K. Reddy and Yogesh M Joshi[*]

Department of Chemical Engineering, Indian Institute of Technology Kanpur,

Kanpur 208016, INDIA.

[*] Corresponding Author, E-Mail: joshi@iitk.ac.in.

Tel: 0091 512 259 7993, Fax: 0091 512 259 0104

**Abstract**

In this work, we investigate the ageing behavior of soft glassy solids of aqueous suspension of laponite under shear flow. We observe that when an imposed time is normalized by a dominating relaxation time of the system, the rheological response at different ages shows superposition. Analysis of this behavior suggests that the structural evolution with age under a deformation field, as represented by the dependence of dominant relaxation mode on age, becomes weaker as the system becomes progressively less homogeneous due to enhanced attractive interactions caused by addition of salt. Creep-recovery behavior at same elastic modulus shows more viscous dissipation for a system having more salt, demonstrating increase in the mechanical fragility. This study shows that an increase in the concentration of salt, which enhances attractive interactions and causes greater inhomogeneity, leads to a state wherein the particles are held together by weaker interactions. This work leads to important insights into how microstructure affects the ageing dynamics. We discuss the observed behavior in the context of ageing in colloidal glasses and gels of aqueous suspension of laponite.

## I. Introduction

Colloidal suspensions near jamming transition demonstrate variety of microstructures depending upon the nature of interactions between its constituent entities. Lately two such important out of equilibrium states of these materials, namely a glassy state and a gel state, have attracted significant attention.[1-21] Non-ergodic character of such states induces ageing wherein the colloidal particles undergo microscopic dynamics in a caged state and lower the potential energy.[22] This dynamics is strongly dependent on the microstructure of the non-ergodic state (or the nature of a cage). Application of the stress field increases energy of the trapped particles and has a profound effect on the ageing behavior.[23-31] In this work, we use an aqueous suspension of laponite to investigate how ageing under the stress field is affected by a microstructural state of the system. We observed that as the microstate of the system became progressively less homogeneous due to enhanced attractive interactions; an evolution of the dominating relaxation mode with respect to age became slower. In addition, a response to the mechanical stress became more fragile with reduction in the homogeneity of the system.

Laponite constitutes disc shaped nanoparticles with a diameter of 25 nm and a layer thickness of 1 nm. The face of a laponite disc is negatively charged while the edge is positive or negative depending upon pH of the system.[32] At pH 10, an electrostatic screening length associated with the face is 30 nm,[11] however there is no consensus regarding the nature of the edge charge at this pH.[3,33] Irrespective of its nature, the magnitude of the edge charge is weaker than that of a face, which leads to a strong repulsion among the laponite particles causing ergodicity breaking at the low concentration of laponite.[11,34] Certainly, we cannot, however rule out the possibility of the edge-to-face attractive interactions in the overall repulsive environment.[8,18] Addition of sodium chloride increases concentration of cations that screen the negative charges on the face enhancing attraction among the particles.[8-10] Thus the microstructure gets strongly affected by the concentration of salt. The phase behavior of laponite suspension with respect to concentration of laponite and that of the salt is extensively debated in the literature, and has lead to different proposals for the phase diagrams.[3,8,10,15,17,18,35] In this work we adopt a view that proposes a homogeneous state for around 1 volume % laponite with no salt which becomes progressively inhomogeneous with increase in the concentration of salt.[3,9,10,17,18,36]

Comparatively homogeneous state at 1 volume % of laponite and no salt has been referred to as a glass by many groups in the literature.[3,10,18,19,34,37,38] It is generally believed that the higher concentration of salt leads to a fractal gel state.[8,10,18,19] Generally a glassy state is distinguished from a gel state based on the morphology which determines above what length scale the system is homogeneous. In the case of glasses, it is the particle length-scale. The microstructure corresponding to a glassy state is disordered. On the other hand, a gel is comprised of a fractal network so that the system is homogeneous above the cluster length-scale. Both these states, though structurally different, have a restricted access to the phase space. Every matter has a natural tendency to lower their energy in search of a possible equilibrium state. For the systems in glassy state, this process is extremely sluggish. The ageing dynamics in the molecular glasses, followed by a rapid decrease in the temperature below the glass transition temperature, leads to densification, so that the system lowers its energy and progressive ordering takes place in the material.[39-41] However in the soft glassy systems, since ergodicity breaking occurs at much coarser length scale, the microscopic dynamics at that length scale causes an evolution of the structure over a prolonged period of time lowering the potential energy. In the soft glassy materials, the nature of microstructure also affects the ageing dynamics. Recently Joshi *et al.*[19] proposed that ageing in colloidal glasses and gels is fundamentally different, with the former undergoing a gradual ordering with age.

Under the no flow conditions, the only characteristic time associated with the ageing system, such as a colloidal glass or gel, is its age and therefore the timescale of the material scales proportional to age. However, after the sufficient ageing time, material may progress towards saturation in the ageing dynamics.[40,42] Application of the stress field influences the ageing process whenever the timescale of the flow field is smaller than the timescale of the material.[25] If applied stress is above the yield stress, complete rejuvenation occurs so that ageing stops and material flows like a liquid.[27,29,43] Intermediate stresses retard the ageing processes so that the dependence of timescale of the material on age weakens.[27,29] This dependence can be estimated by carrying out systematic rheological experiments at different ages and by shifting the data appropriately to yield a master curve.[27,29,40,44,45] In this work, we, for the first time investigated the explicit dependence of dominating relaxation time on age (waiting time) of ageing suspension of laponite as a function of its microstructure at various stresses. We carried out creep-recovery experiments at different ages on an aqueous suspension of laponite at various concentrations of salt. Application of creep time-aging time superposition protocol due to Struik,[40] lead to a very powerful procedure that yielded relationship between the dominating relaxation mode and the age (waiting time) for various stress fields. Subsequent recovery experiments enabled determination of viscous dissipation in the creep flow with respect to the microstates.

**II Experimental**

Synthetic hectorite clay, Laponite RD, used in this study was procured from Southern Clay Products, Inc. The white powder of Laponite was dried for 4 hours at 120 °C and mixed vigorously with water at pH 10 which is maintained at predetermined molar concentration of Na$^+$ ions by addition of NaCl. The pH of 10 was maintained by the addition of NaOH to provide chemical stability to the suspension.[46] The suspension was stirred for 15 min. In this work we have employed stress controlled rheometer AR 1000 (Couette geometry, inner diameter 28 mm with gap 1mm) to carry out the creep experiments. After filling the couette cell with the sample, system was kept idle for a definite period of time in order to achieve ergodicity breaking. To ensure the same, at the end of the waiting period, frequency sweep experiments were carried out on the independent samples. We observed that the elastic modulus was independent of frequency while the viscous modulus was very weakly dependent on frequency in the experimentally accessible frequency range. According to Fielding *et al.*[31] this behavior implies system to be in the non-ergodic regime. After the waiting period, large magnitude oscillatory shear stress was applied to the sample that induces large amplitude oscillatory strain in the sample thereby erasing the shear history. In this shear melting step, the suspension yielded and eventually showed a plateau of low viscosity that did not change with time. The shear melting was stopped at this point in time, from which the age of the sample was measured. Subsequently, ageing of the suspension was monitored by applying oscillatory shear stress with amplitude 0.5 Pa and frequency 0.1 Hz. In this step, the complex viscosity increased rapidly with the age. We carried out the ageing experiments on various independent samples until a predetermined age was reached and performed the creep experiments. To avoid evaporation or the possibility of $CO_2$ contamination of the sample in the test cell, the free surface of the suspension was covered with a thin layer of low viscosity silicon oil. In this work we have used five systems having 3.5 weight % (1.45 volume %) laponite and concentration of NaCl ($C_s$) in the range 0.1 mM to 7 mM. Here $C_s$ represents concentration $Na^+$ ions due to externally added salt or NaOH and does not include concentration of counterions of laponite. All the results reported in this paper relate to 20 °C. The comprehensive experimental procedure including various details pertaining to the shear melting and ageing step are given in a recent paper by our group.[19]

**III Results and Discussion**

Aqueous suspension of laponite undergoes ergodicity breaking soon after dispersing laponite powder in water.[34] In such state every laponite particle can be considered to be trapped in an energy well created by the surrounding particles. The nature of resulting energy landscape is affected by uneven charge distribution, anisotropic shape of the particle, presence of salt and concentration of laponite. A particle trapped in an energy well is a very generic representation that applies to various nonergodic morphologies of the system including glasses as well as gels. Ageing in such state involves structural evolution (or rearrangement) thereby particles lower their potential energy by undergoing microscopic dynamics inside the cage. This structural evolution makes the system progressively more elastic. We have carried out creep experiments on the independent samples at various ages that correspond to different values of elastic modulus. In general it was observed that that increase in the age was accompanied by increase in the elastic modulus. Subsequently, in the creep experiments, strain showed weaker enhancement at the larger value of age (waiting time). In our recent paper,[27] we have shown that such creep curves obtained at different ages show a superposition when ratio of momentary compliance $\left(J(t_w+t)=\gamma(t_w+t)/\sigma_0\right)$ and zero time compliance $\left(J(t_w)=1/G(t_w)\right)$ is plotted against $t/t_w^\mu$. Here $t$ is a creep time while $t_w$ is an age (waiting time) at which the creep experiment were started. Since the relaxation time of the system evolves with the creep time, in order to obtain reliable superposition, creep time considered in the superposition was kept significantly smaller than the age (at most 10 % of that of age).[27,40] The term $t_w^\mu$ represents that factor of the dominating relaxation mode which is dependent on the age for a given stress field. Consequently exponent $\mu$ represents rate at which the dominant relaxation mode increases with the age $\left(\mu=d\ln\tau/d\ln t_w\right)$.[27,29,31,40,44,45] Dominant relaxation mode is related to $\alpha$ relaxation mode of the system[31] that represents cage diffusion time scale. Thus, $\mu$ also represents rate of structural evolution. Greater the values of $\mu$ is, faster is the ageing dynamics.

We carried out the above mentioned superposition procedure for systems with $C_s$ in the range 0.1 mM to 7 mM. Figure 1 shows superposition obtained for systems having 1 mM and 7 mM salt concentration (for $\mu$=0.5 and $\mu$=0.25 respectively). It can be seen that the curvature of the master curve at both the concentrations is different, suggesting time-salt concentration superposition is not possible for this system. The curvature is steeper (more deformation) for a system with higher salt concentration. Figure 2 shows values of $\mu$ for which the superposition is obtained plotted against $C_s$. The value of $\mu$ decreased with increase in the concentration of salt. As discussed before, $\mu$ represents rate of structural evolution or the rate of ageing. An increase in the salt concentration, which enhances attraction among the laponite particles, is believed to change the arrested state to a progressively less homogeneous state. Inset in figure 2 shows variation in $\mu$ as a function of creep stress for laponite suspension at two salt concentrations. It can be seen that the value of $\mu$ for 5 mM system was always lower than that for a 0.1 mM system for an explored range of the creep stresses. Experimental observation described in figure 2 thus suggests that irrespective of the value of stress, an evolution of the structure with age becomes weaker with increase in the concentration of salt. We have also plotted the rate of evolution of complex viscosity $\left(d\ln\left|\eta^*\right|/d\ln t_w\right)$ on the right ordinate as function of $C_s$.[19] It can be seen that similar to the behavior of $\mu$, $d\ln|\eta^*|/d\ln t_w$ decreased with increase in $C_s$. However it should be noted that the complex viscosity (or elastic modulus) of the sample immediately after the shear melting step was observed to be larger for higher concentration of salt.[19] Therefore, as the concentration of salt in the laponite suspension is increased, at any given age, complex viscosity of the system was larger, but the rate of increase of complex viscosity (or rate of increase of dominating relaxation mode) was smaller. This suggests that with increase in the concentration of salt, a saturated state of ageing reaches more rapidly. A saturated glassy state, wherein equilibrium has been attainted in a mechanical response, has been observed for polymeric glasses;[42] however we do not observe such absolute saturation in the laponite suspensions. Furthermore, enhanced yielding in a system with reduced homogeniety at a given stress can also lead to decrease in $\mu$. We discuss this aspect in greater details later in the paper.

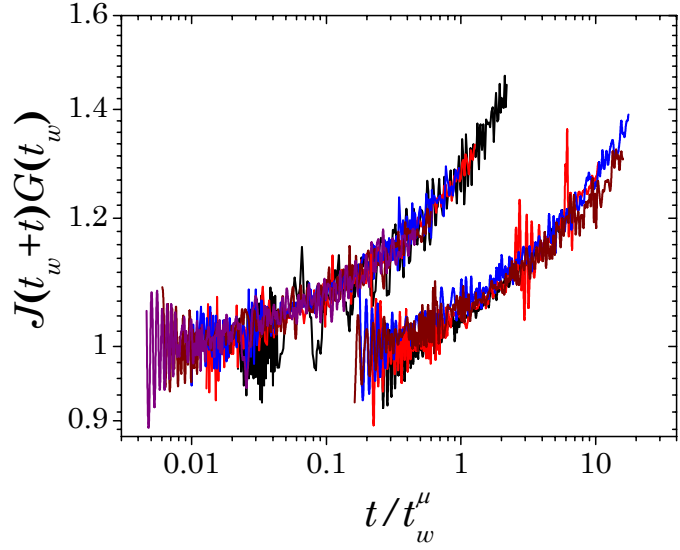

Figure 1. Normalized superposed creep curves obtained at various ages for 3.5 % laponite suspension having 1 mM salt (superposition on the left hand side, $\mu$ =0.5; black 1044 s: red: 2800 s, blue: 4624 s, wine: 11028 s, and purple: 17934 s) and 7 mM salt (superposition on the right hand side, $\mu$ =0.25; black 236 s: red: 495 s, blue: 1054 s and wine: 1505 s). In the creep experiments shear stress of 1.5 Pa was employed. It can be seen that curvature of superposition belonging to a system with 7 mM salt is steeper compared to the one at lower salt concentration.

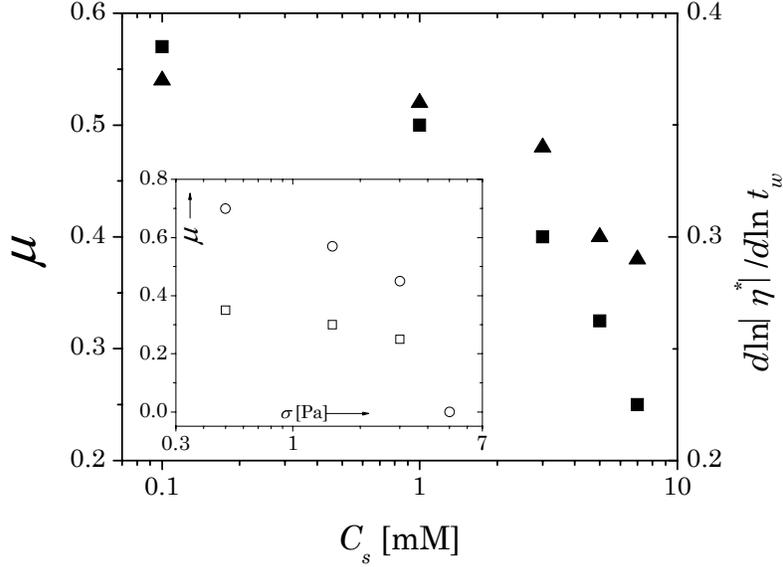

**Figure 2.** Parameter $\mu$ (filled squares) defined as $d\ln\tau/d\ln t_w$ and rate of evolution of complex viscosity $d\ln|\eta^*|/d\ln t_w$ (filled triangles) as a function of concentration of Na$^+$ ions ($C_s$). Oscillatory experiments were carried out at 0.5 Pa stress amplitude while the creep experiments were carried out at 1.5 Pa. $C_s$ =0.1 mM corresponds to salt free system at pH 10. Inset shows $\mu$ as a function of creep stress for $C_s$ =0.1 mM (open circles) and $C_s$ =5 mM (open squares). Irrespective of the value of stress, $\mu$ decreased with $C_s$.

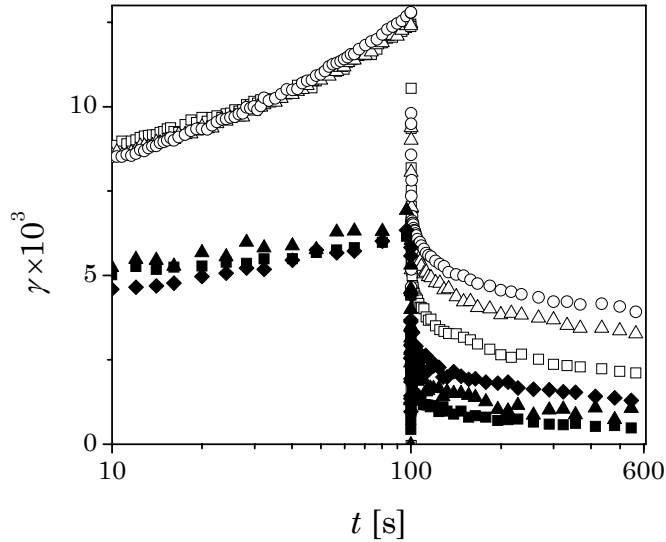

**Figure 3.** Creep-recovery plot for samples having different salt concentrations, $C_s$. (squares: 0.1 mM, triangles: 1 mM, circles: 3 mM and diamonds 5 mM). Open and filled symbols represent the creep experiments that were carried out at the constant value of elastic modulus 191 Pa and 314 Pa respectively. We observed the similar trend in the recovery behavior at the higher values of elastic modulus as well.

Subsequent to the creep experiments, we also carried out the recovery experiments for a period of 500 s. In figure 3, the creep-recovery behavior of samples having a varying degree of salt concentration is shown. The two sets of data represented by open and filled symbols correspond to a constant value of the elastic modulus, 191 Pa and 314 Pa respectively, at which the creep experiments were started. It can be seen that a lesser recovery was observed for a system with more salt but the same elastic modulus. We observed the similar trend in the recovery behavior at the higher values of elastic modulus as well. Lesser recovery represents the viscous flow in the creep response. This result implies that with increase in the concentration of salt, system shows more viscous dissipation in creep.

Microstructural state of the aqueous suspension having 1 volume % of laponite with varying concentration of salt is significantly debated in the literature.[3,8-10,17-19] Particularly at the low ionic concentration (≤ 0.1 mM) of $Na^+$ ions, system is proposed to be more homogeneous. One of the proposals suggests that the ergodicity breaking in this system leads to a repulsive (Wigner) glassy state,[3,9,10,12,37,38,47,48] while the other proposes a house of card type of morphology in which, although there exists an overall repulsion, attraction between a negative face and a weakly positive edge leads to the ergodicity breaking.[2,8] A recent study by Ruzicka *et al.*[18] used small angle x-ray scattering technique and identified this to be a homogeneous state that is attractive in origin. They termed it to be an attractive glass. Mongondry *et al.*[8] represented this state as a homogeneous gel. Thus, it appears that there is a consensus regarding the microstate of the system with no salt to be homogeneous. However there is disagreement regarding whether to represent it as a glass and whether it is attractive or repulsive in origin. For higher ionic concentration, particularly above $C_s$ =1 mM, it is generally accepted that the ergodicity breaking results in a fractal gel formation. Nicolai and Cocard[7] observed that an decrease in the concentration of salt decreased the correlation length. Correlation length is that length-scale of the system beyond which it is homogeneous. In view of this discussion, we have adopted a view that the state of the system becomes progressively less homogeneous with increase in the concentration of salt due to progressive enhancement in the attractive interactions.[3,4,7,8,10]

Two observations mentioned above, namely the slower evolution of structure and enhanced viscous flow under a deformation field with decrease in homogeneity, are reminiscent of a microscopic structure of the system. Furthermore, figure 1 demonstrated that creep time-concentration of NaCl superposition is not possible for this system due to different curvatures of the master curves. This suggests that the ageing and relaxation mechanisms change with the concentration of salt suggesting change in the microstructure of the suspension. However, irrespective of the microstructure, nonergodic state of the system can be represented by a particle trapped in a potential energy well. In such a state, by virtue of the location and the orientation of these particles, there exists a distribution of energies (wells) possessed by the particles. Application of the deformation field enables the particle to climb the energy well.[31] If the applied stress is larger than the yield stress, all the particles overcome the energy barrier of the well; while for the vanishingly small stresses it continues with the microscopic dynamics similar to that of an unperturbed system.[29] For the intermediate values of stresses, a certain fraction of the particles overcomes the energy barrier and leads to a partial yielding of the sample. A value of $\mu$, that represents the rate of ageing or the structural evolution with age, decreases with an extent of yielding and when complete yielding attenuates ageing, $\mu \to 0$.[27,29] We observe that $\mu$ decreased with an increase in the salt concentration. This suggests that for a same magnitude of applied creep stress, with gradual increase in the attraction among the laponite particles, the extent of yielding might be increasing. An enhanced extent of yielding also implies increase in the viscous dissipation leading to a permanent deformation. We observe this behavior in the creep-recovery response wherein an increase in salt concentration leads to a decrease in the recoverable strain. Thus, the mechanical response of the system at a higher concentration of salt, which is believed to have a lesser homogeneous microstate, is mechanically more fragile at the same concentration of laponite.

    We have already proposed a salt free system to be homogeneous above a particle length-scale so that an individual particle has many nearest neighbors. In this state, the caged particle has predominantly repulsive interactions with its neighbors along with the attractive interaction between an edge and a face of adjacent particles. On the other hand, at higher concentration of salt microstate of the system is not homogeneous at particle length-scale (correlation length is much larger) and consequently an individual particle has fewer nearest neighbors compared to that of a state at lower concentration of salt. In this state the

interactions among the particles are primarily attractive in origin. Assuming that the nearest neighbor interactions are the most dominating, it can be conjectured that progressively lesser homogeneous state has fewer constraints to adjust with, compared to that of a more homogeneous state while undergoing ageing. This might explain rapid progress of a less homogeneous state towards saturation. Besides, the present results also suggest that as microstructure of the system becomes progressively less homogeneous, the particles are held in place by overall weaker interactions. Furthermore, in a more homogeneous state, by virtue of more nearest neighbors and repulsive as well as attractive interactions, distribution of barrier heights of energy wells in which the particles are trapped may be broader compared to a lesser homogeneous state, wherein an individual particle has fewer neighbors with primarily attractive interactions. Our recent study on the same system wherein we investigated ageing under oscillatory stress field also suggested broadening of distribution of potential energy well depths with decrease in the concentration of salt.[49] Such situation will enable partial ageing of the system over a larger range of the stress field in a more homogeneous state compared to that of a less homogeneous state and can lead to the observed behavior. However, since the inter-particle interactions and the homogeneity of the microstructure change gradually with an increase in the salt concentration, the observed behavior also shows continuous variation with the increase in the attractive interactions.

**IV Conclusion**

In this work, we, for the first time investigate the explicit dependence of dominating relaxation time on age of ageing suspension of laponite as a function of concentration of salt at various stresses. Following a well defined experimental protocol comprised of shear melting and ageing, creep-recovery experiments were carried out at different ages and creep stresses. Creep curves obtained at various ages show a superposition when plotted against $t/t_w^\mu$, where $t$ is creep time and $t_w$ is age. An exponent $\mu$, defined as $d\ln\tau/d\ln t_w$, corresponds to an extent of structural evolution with age (ageing). We observe that $\mu$ decreases with increase in the concentration of salt which is known to change the microstructure of the sample to a lesser homogeneous state. This suggests that in soft solids of aqueous laponite suspension, the structural evolution under stress is slower as microstructure becomes less homogeneous. Present experiments further suggest that with an increase in the salt concentration, ageing dynamics of the suspension progresses more rapidly towards a saturated state. Subsequent recovery behavior showed larger viscous deformation at the higher salt concentration implying a lesser homogeneous state to be more fragile at the same concentration of laponite. This suggests a possibility of weaker inter-particle interactions with enhanced attractive interactions due to addition of salt. We believe that various observations made in this paper give an important insight into how microstructure affects the ageing behavior under stress in the aqueous suspension of laponite, a model soft glassy material.

**Acknowledgement:** This work was supported by BRNS young scientist research project awarded by Department of Atomic Energy, Government of India to YMJ.